\documentstyle[aasms4]{article}
\input epsf
\begin{document}

\title{Ages of globular clusters: breaking the age-distance degeneracy
with the luminosity function}
\author{Paolo Padoan\altaffilmark{1,2}}
\affil{Theoretical Astrophysics Center, Juliane Maries Vej 30, DK-2100 Copenhagen, Denmark} 
\affil{Dept. of Astronomy, University of Padova, Vicolo dell' Osservatorio 5, I-
35122, Padova, Italy}

\author{Raul Jimenez\altaffilmark{3}}
\affil{Royal Observatory Edinburgh, Blackford Hill EH9 3HJ, Edinburgh, UK}
\authoremail{raul@roe.ac.uk, padoan@tac.dk}

\begin{abstract}
We extend our previous method to determine globular cluster ages 
using the luminosity function (Jimenez \& Padoan 1996). We show that 
the luminosity function depends on both age and distance modulus and that 
it is possible to distinguish between the two. This method provides 
at the same time independent determinations of distance and age of 
a GC by simply counting the number of stars found inside 
specified luminosity bins. The main uncertainties in other 
traditional methods for determining GCs ages are absent (e.g. mixing length, 
color-$T_{\rm eff}$ calibration, morphology of the color-magnitude diagram 
).

The distance modulus is the biggest uncertainty in determining the age 
of GCs. Here we show that the age can be determined with small uncertainty
for any value of distance modulus using the LF and that the LF allows a
determination of the distance modulus itself. This
is explained by the fact that the luminosity function is affected by
a change in distance-modulus in a way that is different from its
time evolution. 

If GC stellar  counts with statistical errors not larger than $3\%$ are 
available, the age can be determined with an uncertainty of about 0.4 Gyr 
(independent of distance modulus, mixing length and color calibration) and 
the distance modulus with an uncertainty of about 0.04 mag.
\end{abstract}


\section{Introduction}

GCs are still the best cosmological clock for measuring the age 
of the Universe. Since the dating of GCs is totally independent 
of the cosmological model adopted to describe the Universe,  
 GCs serve as a constraint for different cosmology scenarios.

Despite the continuous effort carried out during more than 30 
years to give a precise value for the age of GCs, the uncertainty 
in their age still remains of about 4 Gyr. The problem is particularly 
complicated because age and distance have the same effect 
on the morphology of the main sequence turn off point (MSTO). 

In order to tackle this problem, some alternative methods to 
avoid the influence of the distance modulus have been proposed (Jimenez 
et al. 1996). Despite of this it is still interesting to find more 
precise methods that give both distance modulus and age with 
an accuracy better than 5\% in order to be able to constrain 
cosmological models. 

Until now it has been impossible to do so due to the degeneracy 
distance modulus - age. When the MSTO method is used (both 
isochrone fitting and delta V (Chaboyer, Demarque \& Sarajedini 1996)) 
this degeneracy implies that 
a different distance modulus can be mimic with a different mass 
for the MSTO and therefore a different age. In general this leads 
to an uncertainty in the age of 3 Gyr. On top of this, some 
other uncertainties in the stellar physics input leads to 
another 1 Gyr (at least) in the uncertainty of the age.

The luminosity function seems the most natural observable to try 
to constrain both age and distance modulus in an independent and 
accurate way at the same time. 

The luminosity function is a natural  
clock because the number of stars in a given luminosity bin 
decreases with time since more massive stars 
evolve more rapidly than less massive ones. The fact that small differences 
in stellar masses corresponds to large differences in evolutionary time 
explains the power of our clock, rather than being a source of uncertainty 
in getting GCs ages (as it is in the MSTO method).

The luminosity function is also a natural distance indicator because 
the number of stars in a given luminosity bin depends on the position 
of the bin. This dependence is different from the time dependence as 
it is illustrated by the fact that a time translation of an 
evolutionary track for a given stellar mass does not result in 
an evolutionary track of a star with different mass (see Fig. 1).

In this paper we illustrate how the luminosity function can be used 
to determine both the distance modulus and the age of a given 
GC. The paper is organised as follows: in section 2 we describe 
the luminosity function method, we continue in section 3 with the 
dependence of the luminosity function on distance and age. In section 
4 we discuss the effect of the IMF. We conclude with a general discussion 
and the conclusions. 

\section{The luminosity function method}

Stars of different mass evolve along the main sequence at different
speed: the more massive the faster. This means that the number of 
stars inside a fixed luminosity bin decreases with time. In fact the
stars in the main sequence are evolving towards larger luminosities
with the effect that the rate at which stars leave the fixed
luminosity bin towards larger luminosities is higher than the
rate at which they enter the bin coming from lower luminosities.
Furthermore this effect is stronger for more luminous bins (around  
the main-sequence turn off) than for less luminous ones or
for the red giant branch (RGB), so that the whole shape of the LF is
changing with time, and not only its normalization. 

In other words {\it the ratio between the number of stars in two different
bins in the LF can be used as a clock for GC ages}.

The LF is also sensitive to distance modulus. In fact a change in the 
assumed distance modulus corresponds to a translation of the 
luminosity bins when comparing observations with theory, and the 
number of stars in a luminosity bin depends not only on time, but also
on the luminosity of the bin: the less luminous the richer in stars.

For a determination of both distance and age one needs  
at least three bins in the LF: one for the normalization, one for
the age, and one for the distance. A forth bin (the least 
luminous one) is also very useful in order to check for the completeness 
of the stellar counts.

Therefore {\it the LF method for determining age and distance of GCs
consists in the production of 4-bin theoretical LFs for GCs, to be
compared with their observational counterparts}.  

The number of bins should not be larger than necessary (four) since it
is convenient to keep each bin relatively wide in order to reduce the
statistical errors in the stellar counts that are due to uncertainty
in the photometry and to the stochastic nature of the stellar mass
function. If the LF is sampled with many bins they are necessarily
narrow and with large statistical error bars.

We choose to position the normalization bin at the luminosity of 
the RGB, because of the rather slow
time dependence of the number of stars in that region of the CM diagram
(stars of different masses evolve at different speed, but not as much
as in the upper main sequence). The other three bins are all one 
magnitude wide, which is a good compromise between the necessity
of keeping the fourth bin not too faint and the attempt to reduce 
statistical errors by using large bins. The bins, in units of 
$log(L/L_{\odot})$, are: [$\ge$ 1.51, 1.11], [1.11, 0.71], [0.71, 0.31], and 
[0.31, 0.09]. For a GC with $m-M=15.3$ mag these bins correspond to the 
following visual magnitude intervals: [$\le$ 16.3, 17.3], [17.3, 18.3], 
[18.3, 19.3], and [19.3, 20.3]. 

The first bin is used for the normalization. In principle it is not 
necessary to extend it up to the tip of the RGB, but it is convenient
to keep it as wide as possible to include a large number of stars and
reduce the statistical errors.
The second bin is not very sensitive to time, but is very useful to 
determine the distance modulus; the third bin is the most sensitive age 
indicator, because it is at that luminosity (around the main sequence 
turn-off) that the time evolution most strongly affects the stellar 
counts; the fourth bin is finally used to check for the completeness of the 
stellar counts. In fact we will show below that the LF has 
two fixed points (one in distance modulus evolution and one in time 
evolution) that are more luminous than the fourth bin and
can also be used to check for the completeness of the data.

The procedure to obtain the LF from evolutionary
stellar tracks has been illustrated in Jimenez and Padoan (1996).
The first step is that of producing tracks for different stellar masses.
Some tracks are shown in fig. 2, where they are plotted as luminosity
versus time. When a given time is chosen,
the intersection of that time with the tracks gives the mass-luminosity
(M-L) relation at that time. We then assume a power law stellar mass 
function (IMF) that together with the M-L relation allows us to compute the
LF.

In the next sections we illustrate the dependence of the LF on time,
distance modulus and IMF.

\section{The dependence of the luminosity function on distance and age}

Fig. 3 illustrates how the LF changes with distance modulus. LFs 
are plotted for the following values of distance modulus: 15.0, 
15.1, 15.2, 15.3, and 15.4 mag. The time evolution of the LF is shown
in fig. 4 for 10 different ages between 10 and 20 Gyr (the distance
modulus is 15.3). 

It is apparent that the LF is sensitive to both age and distance modulus.
One interesting feature of the LF is the presence of one fixed point.
The fact that the fixed points in the time evolution and in the distance
modulus evolution are at different magnitudes, V=20.1 and V=19.3 
respectively, is a nice illustration of the fact that the two evolutions  
are different from each other. The fixed points are anyway very useful
constraints when trying to fit the observational LF with he theory 
using the two kinds of plots shown in figs. 3 and 4.

One can see that the bin most sensitive to time is the third one. The 
second bin is almost not affected by age but it is the most sensitive to 
distance modulus. In fact the choice of the LF bins focus just on the
attempt of producing a second bin that is sensitive to distance, while
unaffected by age. This is why the degeneracy age-distance can be
broken so well with the LF method.

We find that the LF is really a very sensitive clock and distance
indicator. In fact the number of stars inside the second bin can change
of about 20\% every 0.1 mag and the number of stars in the third bin 
of about 10\% every Gyr. Since at those relatively large luminosities
stellar counts can be performed with uncertainties of only a few
percent, it is clear that the LF method can be used to measure ages
with uncertainty much less the 1 Gyr and distances with uncertainty
much less that 0.1 mag. This means an improvement in both distance and age
determination compared with all other methods that is probably close to 
one order of magnitude!

The procedure to find the distance modulus and the age for a given 
globular cluster can be summarised in the following steps:

\begin{itemize}
\item[i)] The metallicity as well as the helium content are assumed 
to be known from other methods (e.g. spectroscopy).

\item[ii)] Stellar tracks are computed for a given metallicity in a 
large range of mass to cover the GC luminosity bins -- usually from 0.6 to 
0.9 $M_{\odot}$.

\item[iii)] The luminosity bins described in the text (see this section) are 
drawn on top of the tracks (lum vs. time) and the resulting luminosity function for the 
distance modulus and age is obtained.

\item[iv)] An iterative process for fitting the age and the distance modulus 
is carried out until the best fit is obtained.
\end{itemize}

\section{The effect of the stellar IMF}

One of the advantages of the isochrone fitting method is that it is independent on the IMF. Since our method uses the luminosity function, it should be 
in principle dependent on the IMF.  

In fact we find that a reasonable uncertainty in the IMF slope translates into 
an insignificant uncertainty in the age and distance modulus determination. 
The effect of the IMF slope becomes significant only for a  
magnitude larger than 20 ($m-M=15.3$).

Fig. 5 shows the time evolution of a 5-bin luminosity function with 
slope x=1.0 (Salpeter x=2.35), to be compared with Fig. 4, where x=2.5. It 
is evident that the fifth bin of the LF is affected by the IMF, so that 
if observations complete down to that bin were obtained, the LF could 
be used to measure the IMF. Nevertheless, the effect of the IMF on the 
second and third bins, used for distance and age determinations, is 
insignificant. This is illustrated in Fig. 6, where we show those two bins 
for different ages in the two cases x=1.0 and x=3.0. The age difference 
due to the uncertainty in the IMF slope is about 1\% at the age of 15 Gyr.

\section{Discussion}

The first point to address is why the four-bin luminosity function is such 
a good clock and distance modulus indicator. The answer is not hard to find 
if we look at fig. 1 and also look at the physics. In fig. 1 we have plotted 
two panels with two evolutionary tracks. In the right hand side panel we 
show two stellar evolutionary tracks for two different masses (0.80 $M_{\odot}$ 
(dotted-dash line) and 0.70 $M_{\odot}$ (continuous line)). If stellar evolution were a self-similar problem, it should be possible to make a translation in time and mimic one of the tracks with the other. As we show in the left hand side panel this is impossible. It is impossible to mimic the {\it whole} track for 
a given mass with another track of different mass. The only way we can do 
this is by pieces, i.e. we can mimic the RGB but forgetting the rest of the 
track, and also we can mimic {\it part} of the MS but forgetting the rest.
This is the reason why the four-bin luminosity function is able to break 
the age--distance modulus degeneracy. 

The physics behind this is also easy to understand: in different stellar 
evolution stages the luminosity can be express like a power law of the mass 
with an index that depends on the particular stage of the evolution. Since 
the index is different for different stages (giant branch, sub-giant branch) 
it is impossible to mimic two tracks with different masses using a 
time translation.  

A way to show how the age-distance modulus degeneracy is broken is illustrated 
in fig. 7, that is a plot of the uncertainty in the stellar counts (in percent) 
 for different ages and distance moduli around the two values 12 Gyr and 
$m-M=15.2$. The contour plot is obtained by comparing theoretical LFs 
for different ages and distance moduli with a given theoretical 
LF computed for an age of 12 Gyr and $m-M=15.2$. The 
same could be done in a comparison with a real GC, if data were 
available for a cluster of 12 Gyr of age and $m-M=15.2$. In that case 
the contour that corresponds to the uncertainty in the observational 
LF would be chosen, and its extend would give the uncertainty around the best fit.

Fig. 7 shows that, if stellar counts are provided with uncertainty not larger 
than 4\% in each bin (including both the error due to the photometry 
and the error due to the stochastic nature of the stellar IMF), the age 
of the GC could be determined with an error of about 0.5 Gyr {\it 
independently of distance modulus}, and the distance modulus with an error 
of about 0.05 mag, {\it independently of age}.

In the GC age game there are two cards to play: the first one is how to get 
the mass of the stars in the GC from the observables and the second how 
to assign this mass to a given age. The luminosity function method 
addresses the first problem. The second one lies in the field of 
stellar evolution physics and how well understood stellar evolution is.
We now discuss what are the uncertainties in the stellar models used here 
and how this could affect the predicted values for the age and the distance 
modulus.

The stellar evolutionary tracks have been computed with the  latest 
version of James MacDonald stellar evolution code (JMSTAR9). The code 
includes the latest advances in radiative opacities, a very detailed 
discussion of the code can be found in Jimenez \& MacDonald (1996). 
These tracks have been computed with solar scaled abundances and with 
no helium diffusion. It means that in case helium diffusion is important, 
the obtained age of the GC should be shorten in 0.5 - 1.0 Gyr. On the other 
hand, the use of solar scaled abundances is obviously wrong since there 
is large observational evidence that the $\alpha$-elements are 
enhanced in the galactic halo to respect the solar values. 
Nevertheless, this does not mean that our tracks are wrong. The effect of 
enhancing the $\alpha$-elements can be very well mimic using scaled 
solar metallicities (Salaris, Chieffi \& Straniero 1992), 
or just through the effect of 
the oxygen in the CNO cycle (Vandenberg 1992). In the later case it would 
mean an age reduction of about 2 Gyr. 

Although it is not clear yet if helium diffusion is really taking place 
in low mass stars, it is demonstrated that the $\alpha$-elements are 
not in the solar abundance. We therefore bear in mind that our tracks should 
include this but until a new set of tracks is computed we think that a good 
approximation is to follow the above approximations by Salaris and collaborators and 
Vandenberg.

\section{Conclusions}

In this paper we have addressed the question of how to compute GC ages 
using the luminosity function. We have also shown how it is possible 
to break the age -- distance modulus degeneracy. The main conclusions of 
our work are the following:

\begin{itemize}
\item A four bin luminosity function provides a new method to break the 
degeneracy age--distance modulus in GCs.      

\item The same four bin LF allows the determination of age with 
uncertainty of about 0.5 Gyr and distance modulus with uncertainty of about 0.05 mag, if observations are available such that the uncertainty in the stellar 
counts is not larger than 4\% (which can be obtained with modern telescopes 
like HST and NTT).  
\end{itemize} 

\acknowledgements
We thank John Peacock for helpful and constructive discussions during 
the elaboration of this paper. RJ thanks TAC for its hospitality.  PP enjoyed 
the hospitality of the Royal Observatory in Edinburgh were part of 
this
work was carried out.
This work has been partly supported by the Danish National Research Foundation
through its support for the establishment of the Theoretical Astrophysics Center
.

\newpage
\begin{figure}
\centering
\leavevmode
\epsfxsize=1.0
\columnwidth
\epsfbox{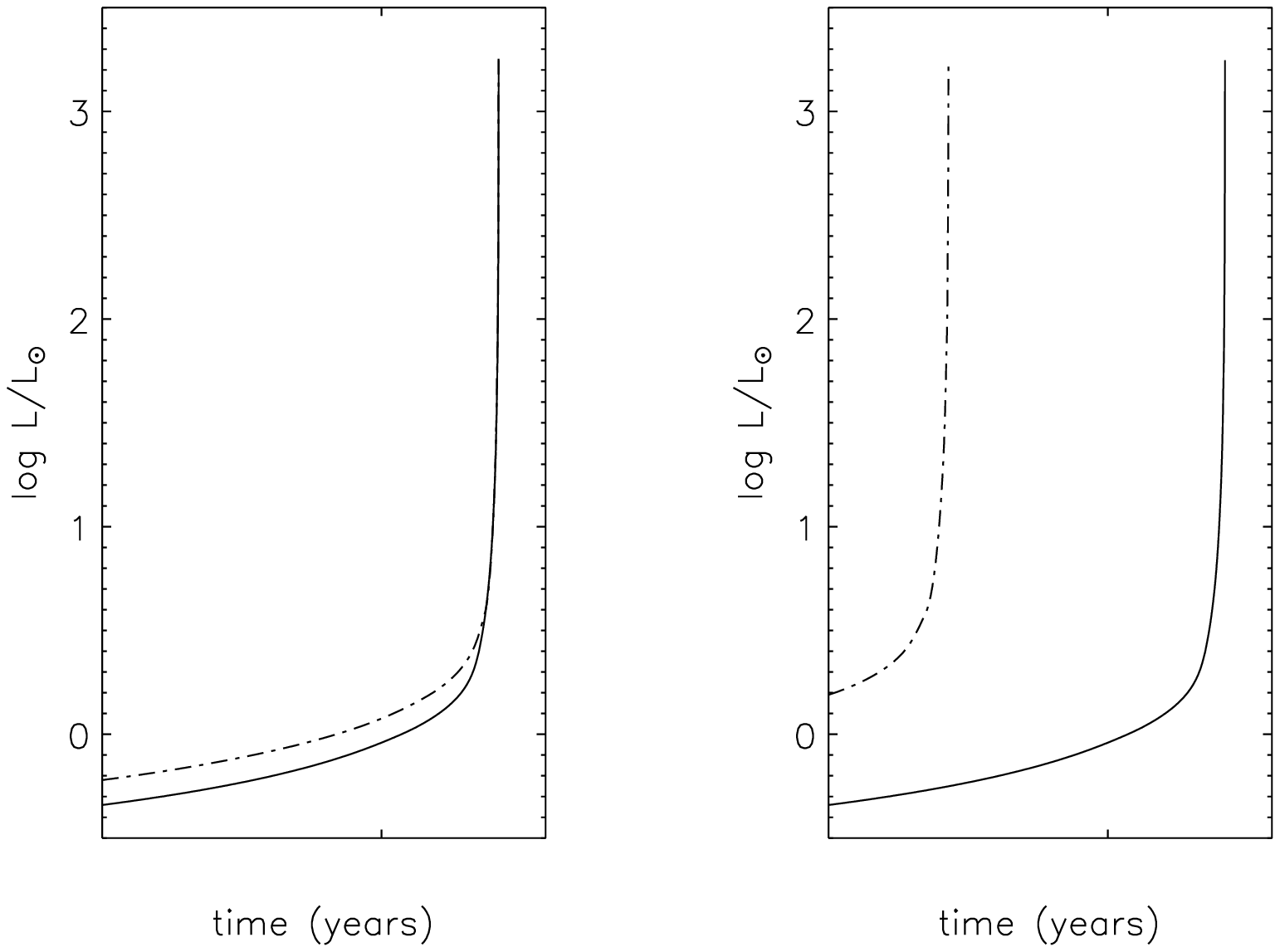}
\caption[]{On the right panel we show the luminosity-time diagram for two 
stellar masses (0.70 $M_{\odot}$ for the continuous line and 0.80 $M_{\odot}$ 
for the dot-dashed line). On the left panel we have performed a time 
translation to try to mimic the 0.70 track with the 0.80 one. It transpires 
from the figure that it is impossible to do so and that it is justified 
to use a four-bin luminosity function to determine the age and the distance 
modulus for a given GC.}
\end{figure}
\newpage
\begin{figure}
\centering
\leavevmode
\epsfxsize=1.0
\columnwidth
\epsfbox{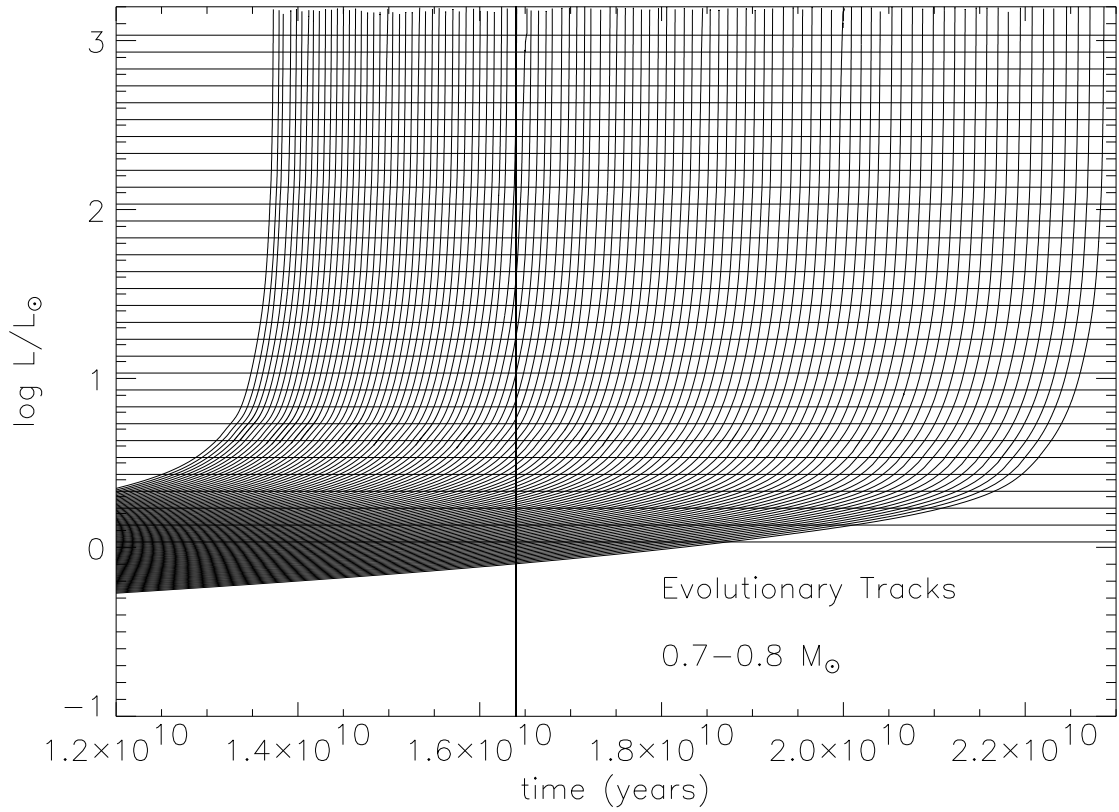}
\caption[]{Evolutionary tracks for stars in the range of masses 
0.7-0.8 M$_{\odot}$.
 The tracks are spaced by 0.001 M$_{\odot}$.}
\end{figure}
\newpage
\begin{figure}
\centering
\leavevmode
\epsfxsize=1.0
\columnwidth
\epsfbox{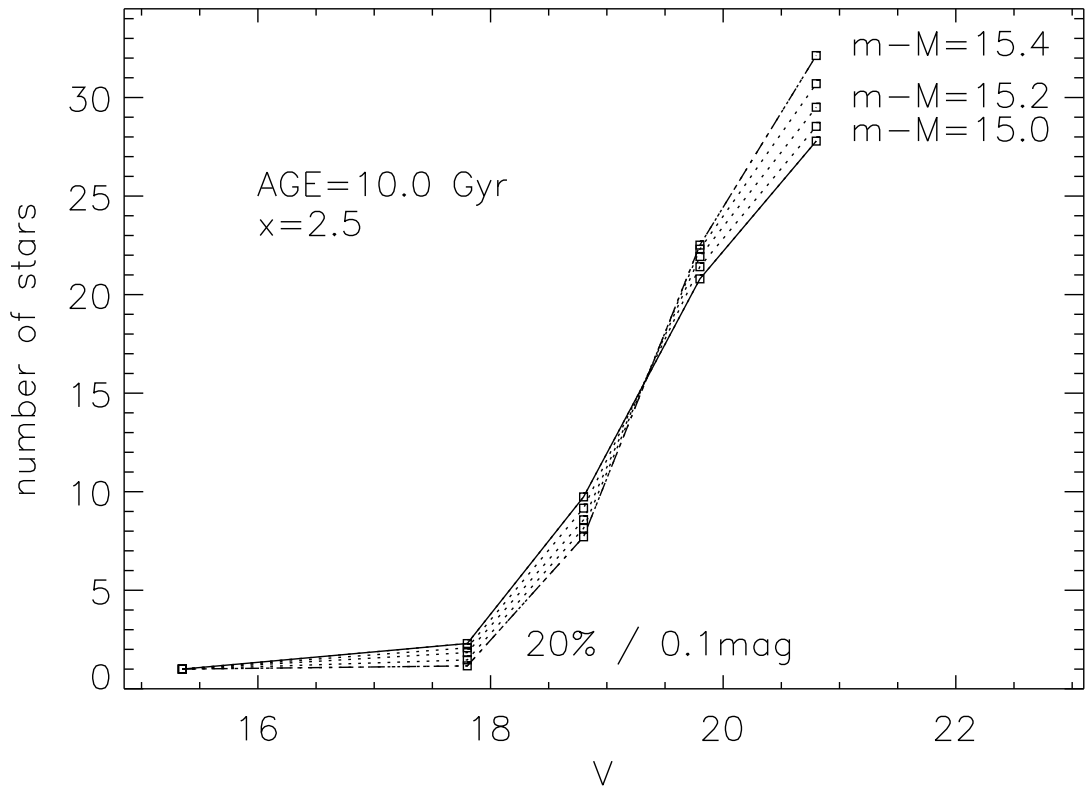}
\caption[]{The four-bin luminosity function used to determine the 
distance modulus. Notice the fixed point in it and the spread in the 
second bin that allows the determination of the distance modulus. In the 
second bin the spread for different distance moduli is of 20\%. }
\end{figure}
\newpage
\begin{figure}
\centering
\leavevmode
\epsfxsize=1.0
\columnwidth
\epsfbox{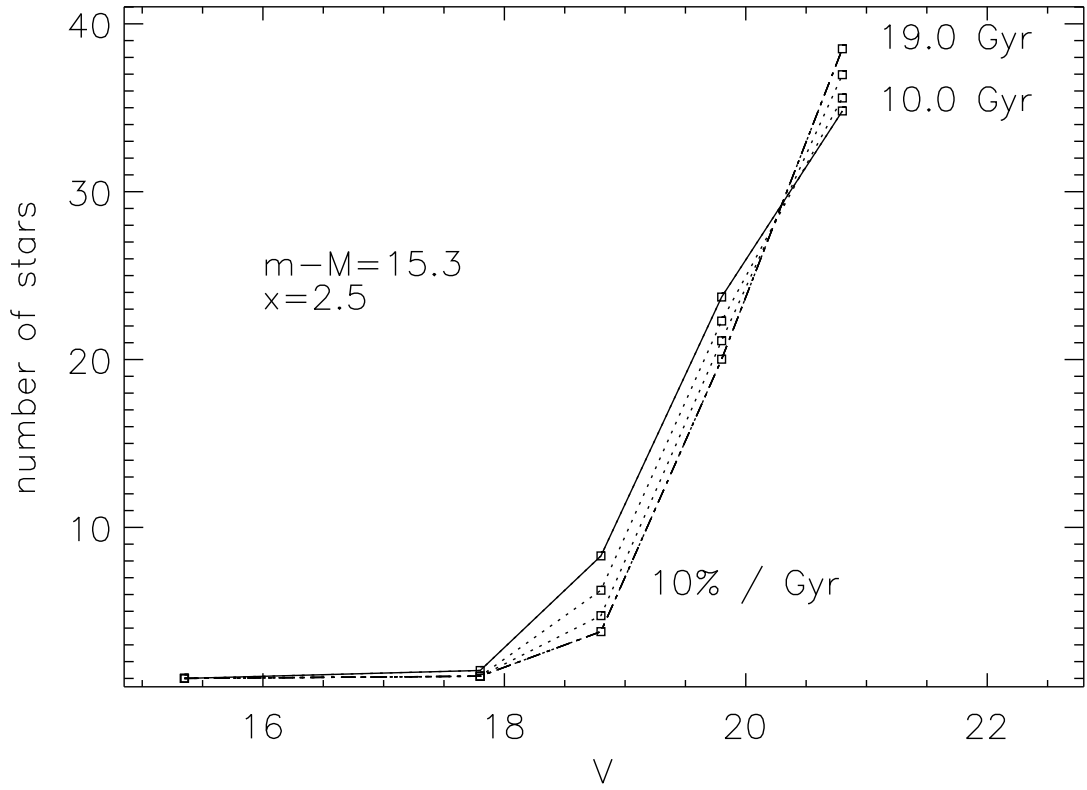}
\caption[]{In this figure we illustrate how to use the four-bin luminosity 
function to determine the age of the GC. The third bin is used to 
do so and the spread among different ages is of 10\% per Gyr, this means that 
if stellar counts are performed with 5\% uncertainty the age can be 
determined with a precision of 0.5 Gyr. }
\end{figure}
\newpage
\begin{figure}
\centering
\leavevmode
\epsfxsize=1.0
\columnwidth
\epsfbox{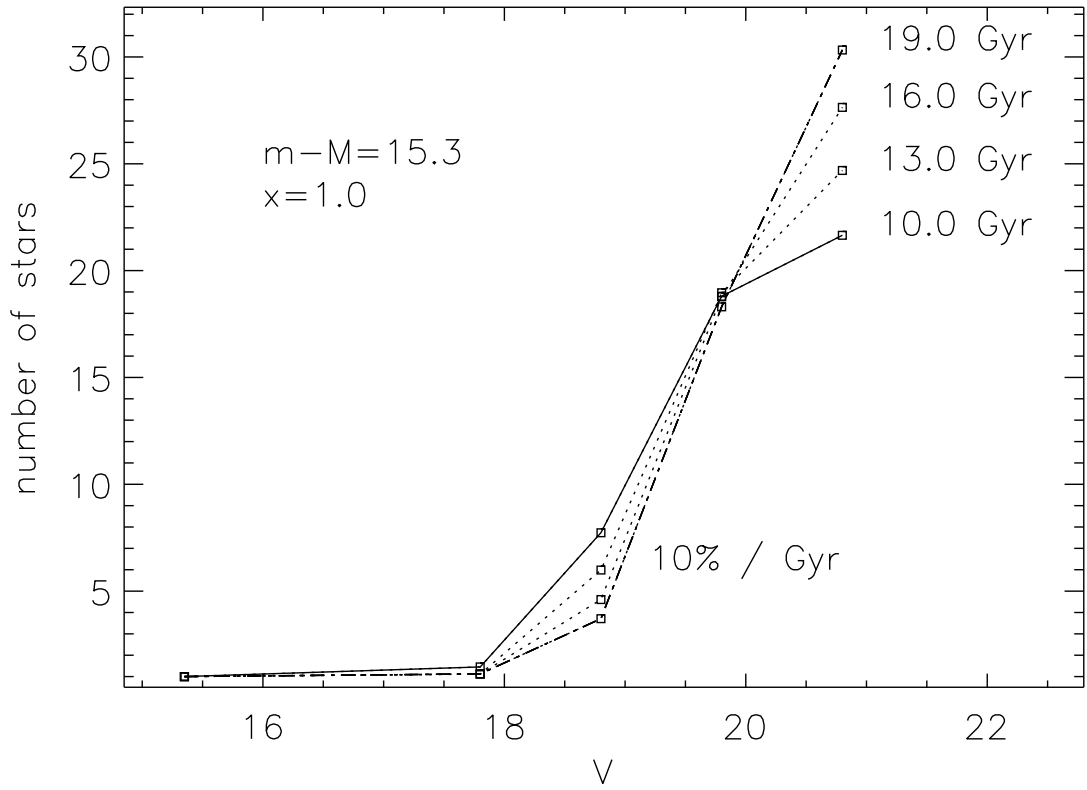}
\caption[]{The effect of a different choice for the slope of the IMF. A 
comparison with Fig. 3 shows that even the choice of such a radically different
 slope for the IMF does not affect in more than 2\% the computed age. }
\end{figure}
\newpage
\begin{figure}
\centering
\leavevmode
\epsfxsize=1.0
\columnwidth
\epsfbox{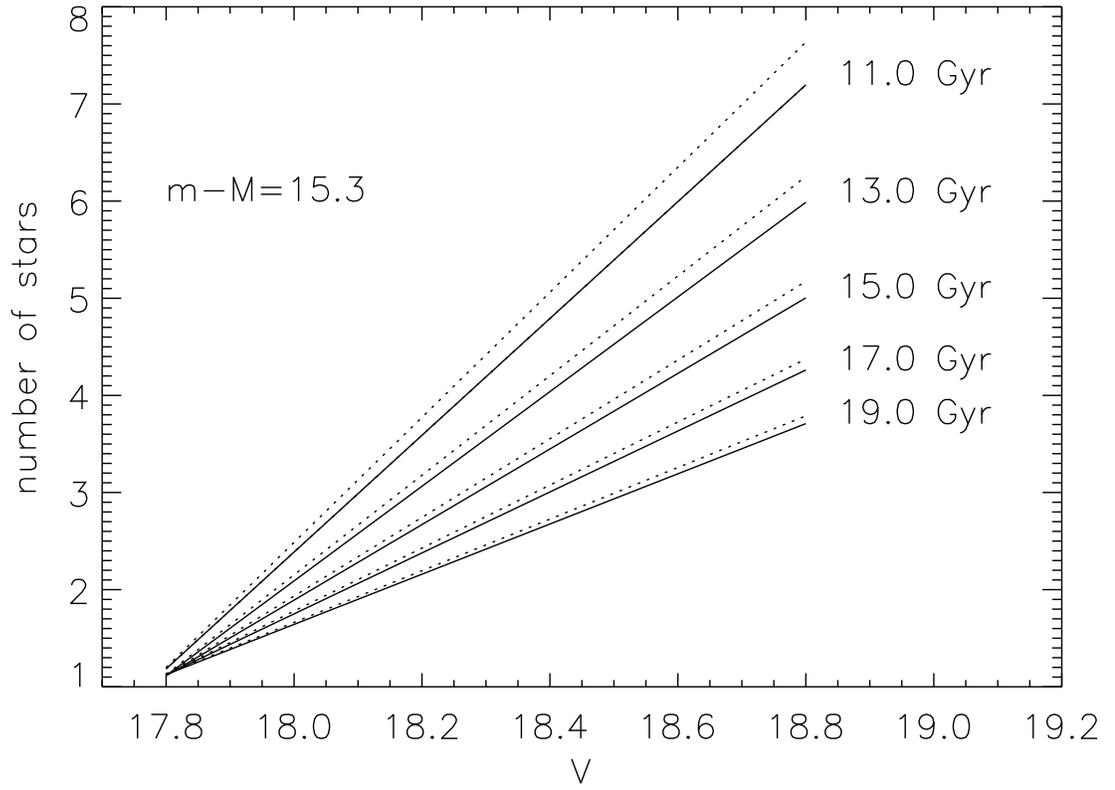}
\caption[]{The effect of the stellar IMF in the determination of the age 
for a given GC using the LF method.  
The dotted lines are for $x=3.0$, the continuous ones for $x=1.0$ (Salpeter 
$x=2.35$). If the uncertainty in the IMF slope is in the range $x=1-2$, then 
the uncertainty in age is only 1\%.}
\end{figure}
\newpage
\begin{figure}
\centering
\leavevmode
\epsfxsize=1.0
\columnwidth
\epsfbox{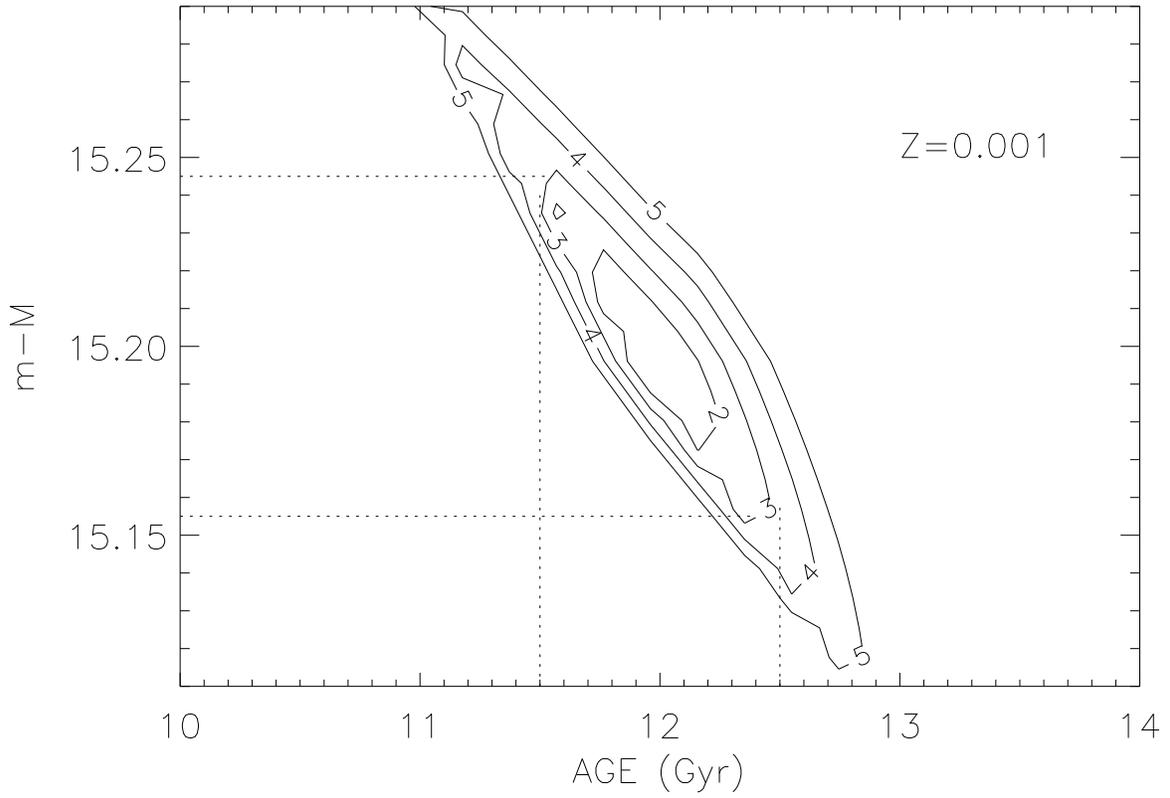}
\caption[]{The contour plot shows the confident levels for the fit to the
age and distance modulus. As expected the contours close around a value and
do not elongate like in the isochrone fitting method. For a synthetic cluster
of 12 Gyr. and $m-M=15.20$ and assuming an error in the stellar counts
of 4\%, the age determination is about 0.5 Gyr and the distance modulus
is determined in 0.05 mag.}
\end{figure}

\end{document}